\documentclass{article}

\usepackage[preprint]{neurips_2023}

\usepackage[utf8]{inputenc} 
\usepackage[T1]{fontenc}    
\usepackage{hyperref}       
\usepackage{url}            
\usepackage{booktabs}       
\usepackage{amsfonts}       
\usepackage{nicefrac}       
\usepackage{microtype}      
\usepackage{xcolor}         
\usepackage{csquotes}
\usepackage{graphicx}
\usepackage{amsmath} 
\usepackage{algorithm}
\usepackage{algpseudocode}

\title{A Foundational Theory for Decentralized Sensory Learning}

\author{
Linus M\aa rtensson \\
IntuiCell AB\\
\texttt{linus@intuicell.com} \\
\And
Jonas M.D. Enander \\
IntuiCell AB \\
\texttt{jonas@intuicell.com} \\
\And
Udaya B. Rongala \\
IntuiCell AB \\
\texttt{udaya@intuicell.com} \\
\And
Henrik J\"orntell \\
IntuiCell AB \& Lund University \\
\texttt{henrik@intuicell.com} \\
\texttt{henrik.jorntell@med.lu.se} \\
}

\begin{document}

\maketitle

\begin{abstract}
In both neuroscience and artificial intelligence, popular functional frameworks and neural network formulations operate by making use of extrinsic error measurements and global learning algorithms. Through a set of conjectures based on evolutionary insights on the origin of cellular adaptive mechanisms, we reinterpret the core meaning of sensory signals to allow the brain to be interpreted as a negative feedback control system, and show how this could lead to local learning algorithms without the need for global error correction metrics. Thereby, a sufficiently good minima in sensory activity can be the complete reward signal of the network, as well as being both necessary and sufficient for biological learning to arise. We show that this method of learning was likely already present in the earliest unicellular life forms on earth. We show evidence that the same principle holds and scales to multicellular organisms where it in addition can lead to division of labour between cells. Available evidence shows that the evolution of the nervous system likely was an adaptation to more effectively communicate intercellular signals to support such division of labour. We therefore propose that the same learning principle that evolved already in the earliest unicellular life forms, i.e. negative feedback control of externally and internally generated sensor signals, has simply been scaled up to become a fundament of the learning we see in biological brains today. We illustrate diverse biological settings, from the earliest unicellular organisms to humans, where this operational principle appears to be a plausible interpretation of the meaning of sensor signals in biology, how this relates to current neuroscientific theories and findings, and how it can be applied to solve body control.
\end{abstract}

\section{Introduction}

\subsection{An evolutionary perspective on sensory input}
In neuroscience, learning and adaptation is a common underlying research topic, yet the question of how learning came to arise gradually through evolution in the first place is rarely raised. However, the answer to that exact question is crucial to understand what constitutes relevant information for the brain. The answer would potentially also be of significance to associated academic disciplines where learning and adaptability are fundamental properties, such as artificial intelligence. Artificial intelligence indeed has strong historical connections to neuroscience~\citep{McCulloch1943-cu, Hebb1962-fg}, but where neuroscience failed to provide answers engineering have prevailed and provided alternative non-biological solutions~\citep{Rumelhart1986-fk}.

In this paper we present a foundational theory that enables biological learning to be explained from the ground up as a consequence of a single simple presumption. \textbf{Individual computational units, such as a biological cell, strives to resolve problems through minimization of the signals received through its available sensors in the form of negative feedback control.} The importance of resolving sensed problems scales to more complex or multicellular structures, or organisms, by utilizing downstream cells as remote enactors of required solutions. We define a sensor to be either internal, reflective of the cell's internal homeostasis, or external, i.e. a membrane receptor or other membrane interfacing structures through which the cell can sense its environment. From this point of view homeostasis is then not just a principal feature of life, but essential to each and every sensor and actuator. We present this theory through a number of conjectures which enables learning mechanisms that do not require backpropagated learning or otherwise global reward signals for a learning system to arise.

To get to this point, we extrapolate from cell biology and evolution that sensed interactions with the outside world in a cell can generally be expressed as threats which need to be resolved. Moreover, cell-internal deviations from the homeostatic equilibrium set points in single-cell organisms are always related to nutritional and biochemical deficiencies in that organism, which are also threats that need to be resolved. This principle is scalable all the way up to explain the meaning of internal signals as mediators of problem signals in multicellular lifeforms, such as plants and animals. 

With this perspective, we argue that the capacity to adapt and learn may be as old as the first life forms on the planet, i.e. present already in some of the earliest unicellular organisms and then inherited as a general cell biological function as species grew more and more complex. We describe how the neuron in principle is not unique from other cell types in this regard, but that neurons simply specialized to facilitate long-range communication of internal signals across cell populations in order to direct the cells towards synchronized actions that could resolve suitable threats and thereby further resolve the origin of each neural signal. Hence, the critical function of neurons became to propagate the problems experienced by other cells, and to adapt their own output such that these problem signals are reduced over time. As a consequence, the functional role of each neuron would be to generate output, for example by way of adapting the receptor count in its synapses (synaptic ‘weights’), such that the neuron can both represent its presynaptic activity in its output and strive to minimize that same presynaptic activity. Given this point of view the synaptic input is considered to be the sensory (or problem) input into the neuron and the act of minimizing that problem, potentially in collaboration with connected neurons, becomes a primary purpose of the brain. As we will gradually arrive at, even the act of when an animal generates movements, thereby causing a multitude of sensory input, is compatible with this fundamental principle of minimizing the animal’s total sensory input over time.

\subsection{Sensory minimization in unicellular and other microorganisms}

Cells are biochemical entities, confined by a membrane, in which each molecule and ion will participate in a huge system of coupled chemical equilibria. Examples of such multi-equilibria systems inside cells can be found in various database tools~\citep{Chang2021-iw} and a comprehensive map has been published as well~\cite{Michal2018-ya}. In most (if not all) cell types, including all animal cells, there is an energy-demanding ion pump, typically the sodium-potassium pump~\cite{Horisberger2004-zn}, which establishes ionic gradients and thereby also an electrical potential across the membrane. The gradients and electrical potential can in turn be utilized as potential energy to power all the different transportation processes across the cell membrane~\citep{Alberts2022-ym, Deamer2008-xo} thereby helping resolve the homeostatic deficiencies within the cell~\cite{Mansy2008-xe}. Since the sodium-potassium pump is highly energy-demanding~\cite{Muangkram2023-cw}, the cells need a constant energy supply to keep them operational. This metabolic catch-22 can be considered the primary problem any biological cell needs to solve, energy is needed to power the machinery that enables energy to be accrued. Naturally, apart from ionic transportation other metabolic reaction pathways are also needed for the energy metabolism, such as converting molecules into formats more easily used by the cell, and for the synthesis of functional molecules, such as the ion pumps and the ion channels. Due to the coupled chemical equilibria, shortage of one molecule type can lead to downstream effects across the molecular cascade reactions hence perturbing those equilibria. Such perturbed equilibria can be considered as internal chemical sensing of that deficiency, which leads to that the problem can be propagated to molecular subsystems more appropriate to mitigate or resolve it.

To exemplify how even the state of the cell’s ion homeostasis is a problem sensor, consider if a cell is put in a solution of ions in which the ion pump cannot operate for example due to a shortage of potassium ions. Because of the constituent leak of potassium out of the cell, which is normally counterbalanced by the sodium-potassium pump, the cell’s membrane potential will gradually depolarize, and all depolarization-sensitive channels~\cite{Catterall1988-ic} will eventually open. The opening of the channels will result in an influx of ions which will in turn cause massive reactions within the cell. For example in the unicellular organism, the influx of calcium across those voltage-sensitive channels can cause activation of protein kinases~\cite{Manning2002-wc, Anamika2008-xq}, which in turn will impact the function of many proteins and enzymes within the cell, altering of internal function or even sometimes engage motility in an attempt to resolve the deficiency. This example can be interpreted as an attempt to use the cell’s local problem solving mechanisms to get the cell to a state and/or an environment where the ion pump function can be restored and the membrane is repolarized, because if that fails the cell will die.

Any cellular metabolic or energy deficiency can in this way be regarded as driving internal sensors that pushes the cell to act in order to restore homeostasis. In a more general sense, in order to select responses, single cell organisms are capable of sensing their own internal deficiencies, detecting or ingesting chemicals through surface membrane receptors, detecting mechanical changes via mechanosensitive ion channels or sometimes even light through photon sensitive internal sensors~\cite{Arendt2008-bj, Brodrick2023-hp, Nilsson2009-sf, Renard2009-ij}. These sensor signals can all be defined as problems; or in the case of external chemical signals, a mediator or a signal taken in across the cell membrane to interact with the internal handling of sensed problems. With this in mind, we posit the following:

\begin{displayquote}
Conjecture 1.1: Sensor signals integrated by a cell constitute problems which that cell needs to resolve.
\end{displayquote}

\begin{figure}[h!]
  \centering
  \includegraphics[width=\textwidth]{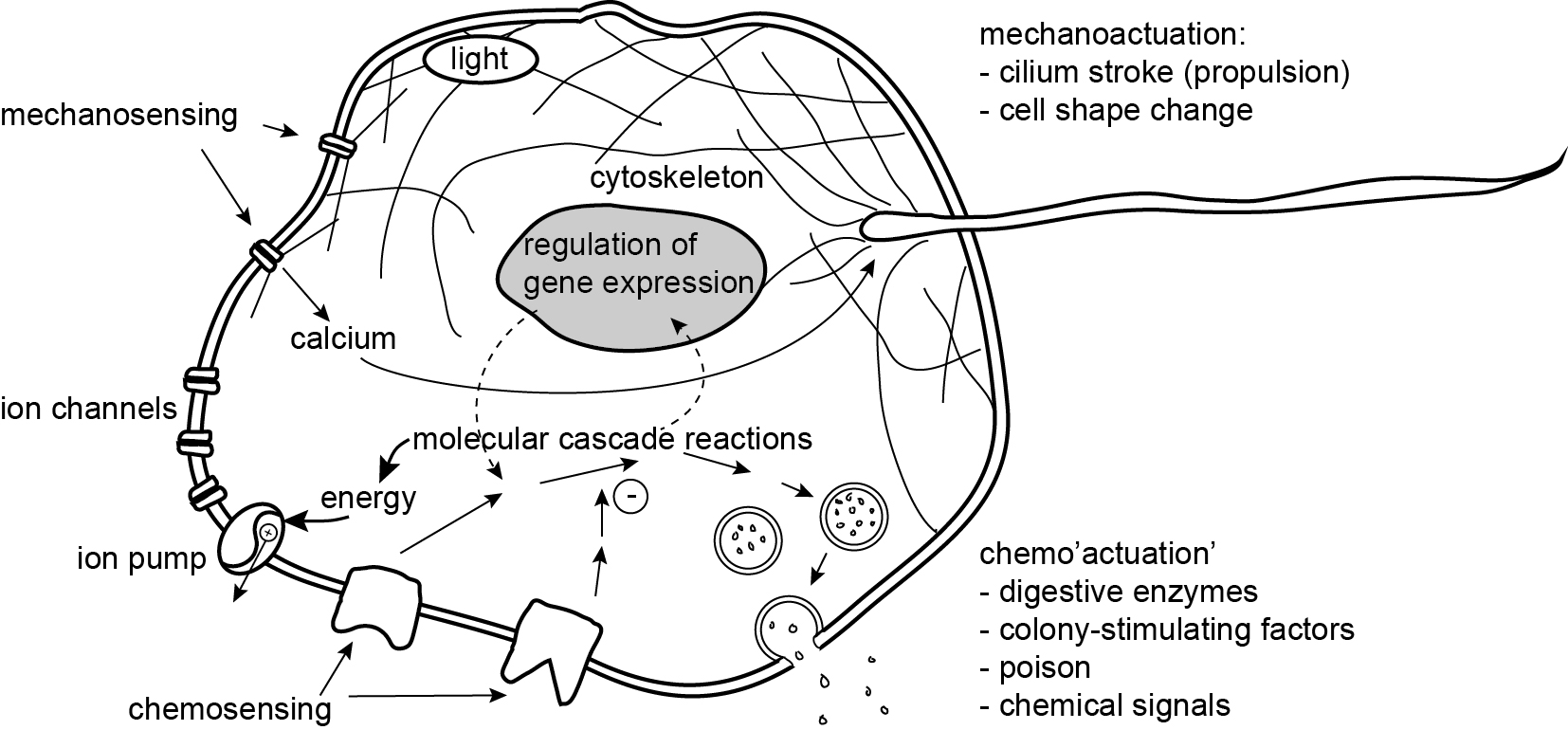}
  \caption{The multifunctional cell. Already the unicellular organism can have multimodal sensing and at least two basic forms of actuation, mechanical and chemical actuation. The primary problem of the cell is its energy need. Energy is required to drive ion pumps, which in turn indirectly drive all transportation across the cell membrane and enable the cell to maintain its internal biochemical equilibrium. Molecules taken in are used in a multitude of metabolic processes, and the cell regulates its gene expression in order to provide the enzymes accelerating those molecular cascade reactions. Chemical sensing is achieved by receptor molecules, typically proteins, and can impact those cascade reactions. A cell can also be capable of transporting chemical signals out across its membrane, for example via exocytosis (‘chemoactuation’). It can also sense mechanical stress in its cytoskeleton, which can be coupled to mechanosensing channels~\cite{Brunet2016-cj}. The latter, when activated, can let in some ions, typically calcium, which can mobilize contraction in the cytoskeleton, but also the cascade reactions. This can lead to cell shape changes, or even propulsion, if the cell in addition has a cilium. Some cells even have light sensors, where the phototransduction into biochemical energy can end up changing for example the calcium level in the cell, thereby driving actuation. These functional components of the cell are known from basic cell biology~\citep{Alberts2022-ym, Arendt2008-bj}.}
  \label{fig1}
\end{figure}

To acquire energy and nutrients, the cell generally needs to navigate or interact with its surrounding environment and the behavioral responses that the cell can exhibit in relation to sensor signals are not limited to the intracellular molecular cascade reactions. These intracellular reactions can also be coupled to external actuation. Either chemically, by releasing or pumping chemicals into the near environment, and/or mechanically, by changes in the cytoskeleton that cause changes in cell shape and in existing cases also in the cilium, which can propel the cell in a fluid or a along a surface~\citep{Alberts2022-ym} (Figure~\ref{fig1}). In principle, the more the cell needs nutrients, the more it should be pushed to act by its internal sensing, which is in line with observations from one of the early multicellular organisms~\cite{Giez2024-ha}.

\begin{figure}
  \centering
  \includegraphics[width=\textwidth]{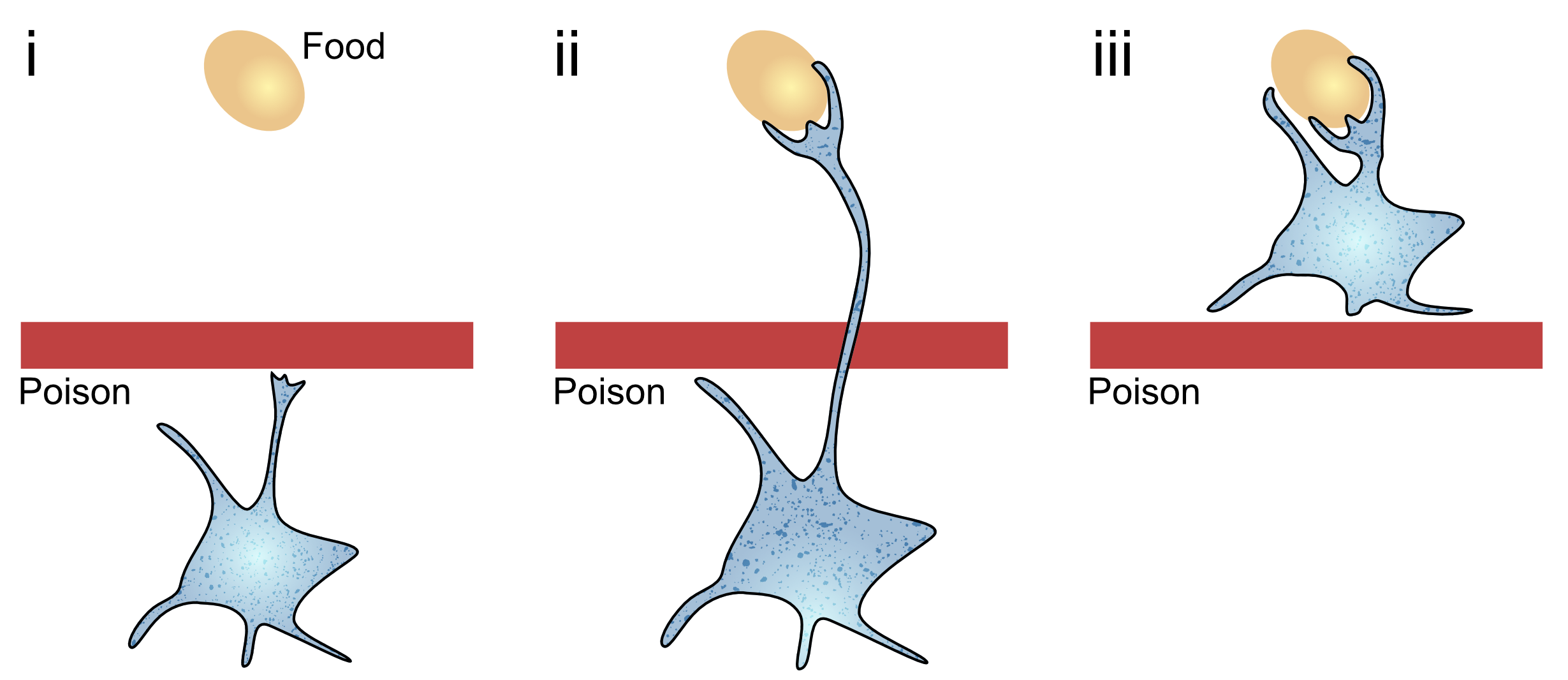}
  \caption{Adaptation of intracellular molecular input-output reactions in unicellular organisms. This is a depiction of an experiment designed to illustrate amoebal behavioral adaptation and behavioral memory. The ‘task’ was to learn to overcome a barrier of a potentially harmful chemical to be able to find nutrients. The amoeba was shown to be able to solve the task where the cell learns to overcome a natural aversive reaction from the chemical in order to solve the greater problem consisting of cell-internal sensing of a lack of nutrients. Interestingly, once the amoeba cell had learned to solve this task, it would solve it much more rapidly the next time it was exposed to the same task and the adaptation was moreover specific to the noxious chemical used in the barrier. This illustrates that the input-output reactions of the intracellular molecular cascade reactions are adaptable and that the behavioral adaptation is stored like a memory. (i) The cell has a nutritional need. This drives activation of the cytoskeleton (Figure~\ref{fig1}) such that the cell changes shape (mechanoactuation) and sends out filopodia that tries to find food. The filopodia hit the chemical barrier, which tends to inhibit its further travel since the cell has membrane surface receptors for this chemical and the activation of those receptors inhibits the extension of the filopodia. (ii) The cell-internal sensing of a lack of nutrients soon overcomes the blocking effect of the barrier and the filopodia instead reaches the food particle. (iii) The whole cell is migrating across the barrier to more rapidly ingest the nutrients. Note that this experiment also illustrates the principle of sensory prioritization by the cell, which we address in this paper. (Figure adapted from Tang and Marshall, 2018~\cite{Tang2018-po}).}
  \label{fig2}
\end{figure}

These responses of the cell are not static but fundamentally adaptable~\cite{Tang2018-po} (Figure~\ref{fig2}). The intracellular molecular cascades triggered by the sensory inputs define the input-output reactions of the cell. These input-output reactions are modifiable through adaptations of the cell’s gene expression~\cite{Flavell2008-yc, Lindberg2010-uh, Ogata2015-ib, Hrvatin2018-wj, Tyssowski2018-mw, Biswas2021-ew}, which in turn alter or modify its molecular reactions, essentially by regulating its internal enzymatic processes for specific biochemical equilibria. Similarly, its expression of sensors (i.e. membrane receptors and ion channels) can be up- or down-regulated by environmental factors to change the gain in those input-output reactions~\cite{Flavell2008-yc, Hrvatin2018-wj, Tyssowski2018-mw, Zhang2003-ms, Moody2005-oi, Pratt2007-lb, Belmeguenai2010-nm, Kuba2010-ff} so that the cascades can couple experienced problems to suitable output responses, even with complex intrinsic functionality within the regulation itself~\cite{ Biswas2021-ew}. We therefore advance the following conjecture:

\begin{displayquote}
Conjecture 1.2: The cell couples problems, i.e. the sensor signals, to responses which trigger some correction aiming to resolve those sensor signals.
\end{displayquote}

To reiterate, the only correction available for some unicellular organisms could be to adapt the internal metabolic process/molecular cascade reactions and pumping specific molecules and ions across the membrane. Other organisms may in addition have the option to correct by releasing or ingesting chemicals, through exocytosis or endocytosis, which in turn can alter the membrane potential or move the cell~\citep{Alberts2022-ym}. In motile unicellular organisms, the correction can also include cell reshaping or movement of the whole cell through ciliary propulsion~\cite{Arendt2008-bj}. These processes can also involve transportation of ions across the membrane, and a normal cell typically has a huge diversity of ion channels and ion transport mechanisms (see for example Podlaski, 2017~\cite{Podlaski2017-kw}). As membrane channels and other surface proteins will be an important part of shaping the cell’s responses to sensory inputs, we posit the following:

\begin{displayquote}
Conjecture 1.3: Each membrane interface in a cell, such as a channel, cilia, or receptor, is evolved either as part of a sensor for some problems, an actuator with which to resolve such problems, or both.
\end{displayquote}

Differential relative expression levels of the genomic set of ion channels and surface membrane receptors can be used by the cell to regulate its input-output reactions~\cite{Flavell2008-yc, Lindberg2010-uh, Ogata2015-ib, Hrvatin2018-wj, Tyssowski2018-mw}. Thereby, cells with the same given set of sensors and actuators can independently adapt themselves to react differently to each sensed quality, and also modulate the response to activation of a second sensor as a response to activation of a first sensor. But important to the presently proposed schema is that a sensed problem must not be ignored or discarded until it has been resolved. Conversely, any change to the internal biochemical input-output reactions, which discards a sensory capacity allowing the cell to perceive and adapt to a problem, will decrease the survival chance of the organism in any suitable environment to trigger that problem and would be negatively selected for in the evolution within that environment. We further reason around the implications of this in the next section where we arrive at conjectures 1.4 and 1.5.

\subsection{Multicellular evolution}

Early unicellular organisms gradually developed into multicellular organisms, Urmetazoans. Multicellular organisms were initially a lump, or a colony, of homogenous cell types that likely formed colonies because a shared ingestion could improve their common survivability by a more efficient resource utilization~\cite{Nielsen2008-eu}. Subsequently, cell type specialization, or cell differentiation, emerged as problems related to digestion and energy metabolism could be solved more efficiently in multicellular organisms that had more specialized cell types. As this happened, it became increasingly complex and in many cases costly for all sensory signals within the multicellular organism to be omnipresent across each differentiated cell in the organism. Instead, this development eventually necessitated transmission of local sensory signals to remote cells more suitable to respond to the detected problem.

\begin{figure}
  \centering
  \includegraphics[width=\textwidth]{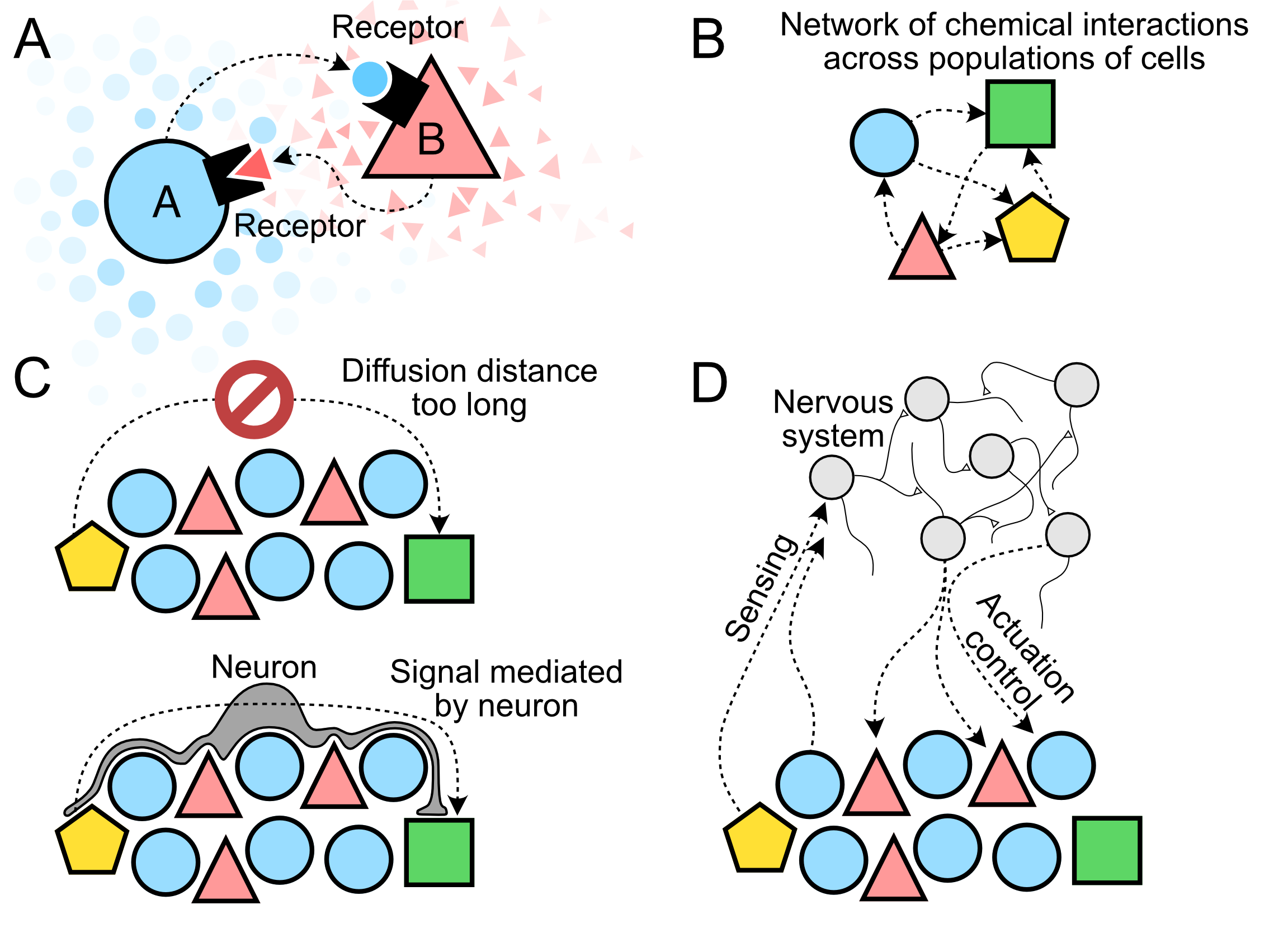}
  \caption{Networks of chemical interactions between cells and the need for neurons. (A) In small multicellular organisms, cells can communicate with other cells via diffusion of signal molecules. If the cells are differentiated, then cell A can only signal to cell B if cell B has a membrane surface receptor for signal A, and vice versa. (B) In organisms with multiple differentiated cell types, cellular interactions can be made highly specific depending on which membrane receptors the specific cells express. Hence, a network of cell-to-cell interactions can be formed. (C) When the multicellular organisms grew larger, relying on diffusion was no longer effective. This necessitated the emergence of specific cell types which were longer than other cells so that they could resolve the problem created by the organism becoming too large for diffusion to be an effective form of communication. Neurons resolved this by developing morphological processes that gradually became axons~\cite{Jekely2021-mo, Jekely2015-rm}. (D) Neurons later developed synapses as an effective means of transmitting signals, and then they could also rapidly communicate between each other, as well as exerting control of actuation in other cells based on the sensor signals received. In this way, the neural network became an entity of its own, while still having the purpose of minimizing the sensor signals arising from the various cells of the body.}
  \label{fig3}
\end{figure}

Therefore, in an emergent strategy, early multicellular organisms built highly complex networks of chemical signaling (Figure~\ref{fig3}A,B), and many of these diverse chemical signals, i.e. neuropeptides and other low-molecular weight compounds, are still used as neurotransmitters by the hypothalamus in animals today~\cite{Arendt2008-bj,Jekely2015-rm}. This is an extension of the same strategy already present within individual cells as discussed above, i.e. where a chemical process triggered by a sensory signal results in an internal chemical problem signal that impacts the molecular cascade reactions within the same cell. In multicellular organisms, those chemical problem signals can in addition be passed externally, between the cells, representing the request of a cell to solve a problem that it does not have the capacity to solve by itself (see Figure~\ref{fig3}). This could naturally happen also in colonies of homogenous cell types. The advantage of multiple differentiated cells is that the set of responses, or problem solutions, that the organism is capable of can be made more diversified and scaled to enact larger changes in the surroundings. Incorporating these points, we propose the following conjecture:

\begin{displayquote}
Conjecture 1.4: Some cells can convert a sensed problem into a format detectable by another sensor. Depending on the method, the converted problem signal can then be detected by the same cell, or by one or more remote cells.
\end{displayquote}

Then, between cells in a multicellular body, each individual cell in the cellular network has to fulfill a contract for the organism to survive, that a signal transmitted through the organism is preserved and emitted until its origin is resolved, e.g. through actuation - at which point it is expected to no longer be a problem. Then we arrive at the following:

\begin{displayquote}
Conjecture 1.5: A problem transmitted through a cell or between cells must not be discarded unless resolved.
\end{displayquote}

In many circumstances, this gives rise to secondary effects: Two, or more, competing problems can through their transmission give rise to an externally observable action that resolves one of the problems and simultaneously a prolonged activity within the organism to retain the others, which in this period may very well grow larger. Such phenomena can give rise to situations where the activity increases in one part of the organism as one problem is retained - or even created - in the process of solving another, more acute problem.

\subsection{Neurons and Networks}
As evolution towards larger organisms and more differentiated cells progressed, diffusion distances started to become too large for efficient transmission of nutrients and signals between organs, necessitating both a circulatory system and efficient conduction of intercellular signals. It has been suggested that around this time, the earliest neurons started to differentiate as elongated cells that grew out long cellular processes in the form of axons that could release chemicals across larger distances~\cite{Jekely2021-mo, Jekely2015-rm} (Figure~\ref{fig3}C). Notably, from our perspective this emergent nervous system led to the situation that more distant cells could receive nonlocal problem signals at almost the same time and with the same magnitude as local cells. In larger lifeforms, the long-range cell-to-cell communication capabilities were further extended by the emergence of a faster electricity-based signaling strategy to quickly transmit sensed problems to remote parts of the body~\cite{Jekely2021-mo, Jekely2015-rm}, which then eventually became myelinated to reach even longer distances as bodies increased in size. If the proposal of the negative feedback control mechanism being of central importance holds true both in individual cells and neural networks composed of multiple cells, it would imply that neurons simply have the purpose of propagating their synaptic inputs through electrical signals in much the same way as the generic cell propagates its sensory signals internally through molecular cascade reactions. The purpose would be to ensure that sensory signals are experienced or retained as unsolved problems and not lost in the transmission between the neurons, and to this end moving presynaptic terminals amongst potential postsynaptic neurons or altering the weights of those synapses, to ensure distinct problems are transmitted to distinct, useful target neurons within the network which can be observed to resolve the initial problems.

The evolution of neural cells, and possibly the divergence between plant-like and animal-like organisms, is also closely linked to the evolution of cells with contractile properties~\cite{Jekely2015-rm, Arendt2021-wr}. As organisms increased in size, more efficient energy intake became necessary to assist the specialized cells within the larger organism, and subsequently more effective means of propulsion as nutrients were depleted faster. The evolution of cells with contractile properties enabled both the speed-up of ingestion due to passage of sea water into the digestive canal, and movement of the organism as a whole towards better nutrient sources, as exemplified by the hydra~\cite{Szymanski2019-eq, Wang2023-bs}. Hence, the multicellular organisms were now provided with another ‘tool’ of actuation, even though contractile cells or muscle cells just represent a further differentiation of the mechanoactuation capability provided by the cytoskeleton present already in the generic cell (Figure~\ref{fig1}), with which the organism could solve its problems of nutrient supply.

Whereas early neuronal networks like the hydra were distributed neuronal nets with only minor regional specialization, evolution gradually gave rise to nervous systems with anatomically co-localized sets of neurons (Figure~\ref{fig3}D). Examples of well-known anatomical structures of such co-localized sets of neurons in vertebrates are the cerebellar cortex, the neocortex, the superior colliculus and the spinal cord. Such structures were gradually also enclosed in specialized anatomical structures such as the cranium and the vertebral column. Neurons of such co-localized sets would have similar synaptological distances from the different available sources of sensory information. This means that the signals received by a co-localized neuronal set would have passed through the same number of other preceding co-localized sets of neurons, each of which would have attempted to solve sensed problems to the best of their capability. This fundamentally simple hierarchy would determine in what way downstream sets of neurons could contribute to yet unsolved problems, and would intuitively describe why different co-localized sets of essentially similar neurons in contemporary animals are often perceived to have specific functions. In view of the this, we formulate the following two conjectures:

\begin{displayquote}
Conjecture 1.6: Foundational brain structures are not extracting information to evaluate and learn from, but communicating sensed problems efficiently in a common form from cell to cell, so that each detected problem is separated and resolved by a suitable actuation mechanism.
\end{displayquote}

\begin{displayquote}
Conjecture 1.7: States which arise in the brain are byproducts of the process described in conjecture 1.6.
\end{displayquote}

In more complex animals, the number and diversity of the sensors that are distributed across all tissues and anatomical parts of both the musculoskeletal apparatus and the internal organ systems is striking~\cite{Wang2024-or}. It could be argued that the animal which can coordinate the needs of these individual sensory modalities in the most efficient manner will have a competitive edge in the evolutionary race. Over time, solving this coordination can lead to a network that fulfills the need for survival in an effective way given the constraints of the environment and of the chemical and mechanical properties of the body itself.

To achieve this coordinated network, where each cell or neuron solves or propagates its locally experienced problems, can be seen as a process of sensory minimization which is the central principle presented here. That simple principle makes it easier to explain how the embryonic brain can ‘bootstrap’ or self-organize its own activity ~\cite{Enander2022-py, Enander2022-ht} against what is essentially an unknown landscape of sensory properties, while also maintaining stable levels of network activity over time. It could also be important to make evolution work at this scale of life. Each cell or neuron solves or propagates its unresolved problems relatively independently in a common (electric) form, which makes it easier to add components, such as more cells, or cells with a new type of differentiation, i.e. adaptations with respect to the cells’ internal biochemical input-output functions and/or sensor sets. Moreover, evolving anatomical features, such as when a limb gradually becomes a fin or a wing, does not need to be coupled to a corresponding evolutionary adaptation in the controlling neural circuitry, which would otherwise make it very hard for evolution of new species to occur at all. If the cells of the neural system are instead designed to solve the central survival problem using whatever actuation ‘tools’ they have at their disposal~\cite{Enander2022-py, Enander2022-ht} differential evolution becomes possible.

\subsection{Sensory Prioritization and Negative Feedback in organisms and animals}
Whereas it is hard to prove the conjectures proposed throughout this article, in this section we will present diverse examples indicative of the central presence of the sensory input minimization principle through negative feedback control.

One example can be found among the germ cells of corals (planula)~\cite{Brodrick2023-hp}. Light sensitive cell compartments can impact the local calcium levels in the cells, which is a sensory signal that impacts the cell’s propulsive activity. The coral planula needs to avoid getting stuck in the parts of the seafloor that are in sunlight, and the light sensors and mechanosensors are used to control the propulsion to this end. Once the light sensor activity has gone down, the germ cell has found a more advantageous spot to plant a new coral animal. In this case, it is clear that the sensory input to the germ cell is definable as a ‘problem’, which the cell needs to solve to the best of its capability given its limited resources to act.

Primitive multicellular organisms seem to have developed a variety of mechanosensors early in evolution where the primary purpose has been to improve the protection of the organism against external threats~\cite{Renard2009-ij}. In larger animals, the number of mechanosensors and chemosensors can easily be in the range of 100,000’s, distributed internally, in the inner organs~\cite{Wang2024-or} and externally, in the movement apparatus and skin.

Regardless if the organism has 10 or 10,000 sensors, a sensor signal can be considered to be acute if it is imminently critical for survival or system integrity. In more complex animals, we can for example point at the carbon dioxide buildup signals which identify imminent risk of asphyxia, or chemosensors detecting actual tissue damage in skin, muscles or internal organs, where tissue damage can arise from physical interaction with the external world or from internal metabolic issues.

Sensors of lesser priority, but which may still dominate brain activity, can be exemplified by the mechanosensors in the body that can indicate, for example, insufficient muscle tension (group (gr.) Ia afferents), overextended limbs (group II), or overloaded tendons (group Ib)~\cite{Enander2022-py}, or by the retina’s sensitivity to movement in the periphery of its visual field~\cite{McKee1984-pq}. The better the lower-order networks like the spinal cord and the oculomotor system are working, the more effective they will be at maintaining low levels of activity for these sensors reflexively, enabling the activations of the higher-order networks to be dominated by higher levels of activity originating elsewhere. Should one of those lower-order networks fail to assist in that control however, the given sensor activity could rapidly dominate brain activity and require intervention or risk escalating in the form of potential tissue damage. This dynamic hierarchy of control is covered by conjectures 1.6 and 1.7 specifically.

Indeed, the sensory afferent systems of the skin, ligaments and muscles are arranged in terms of falling activation thresholds, where the sensors of the highest mechanical threshold are chemosensors (group IV muscle afferents~\cite{Amann2015-et}, C fiber afferents in the skin~\cite{Mense1996-al}) signifying that the mechanical load, for example in the skin, has become so large as to tear the cells of the local tissue apart. In fact, even the skin sensors of the lowest mechanical threshold are in essence shear force sensors~\cite{Hayward2014-za, Jorntell2014-jn, Rongala2024-ad}, implying that every sensor type in the skin to some extent forebodes tissue tear but to different hierarchical degrees, in a way that can support a sensory prioritization schema by the nervous system.

A similar interpretation can be applied to the different types of mechanosensors in the muscles. A prediction that can be made from this schema is that for a well-adapted brain, the likelihood of activation of a sensor type is inversely related to its activation threshold, i.e. for example in a given muscle the group II afferents would be less likely to be activated than the group Ia afferents (experimentally support from Loeb et al, 1985~\cite{Loeb1985-uu}) whereas group IV activation should be rare.

However, under specific circumstances, the default sensory prioritization can quickly change as we for example temporarily cease to breathe when lifting heavy objects, or if we in sports or combat incur a body damage but do not notice it until after the acute stressful situation, such as the competition or training session, is over.

What we suggest and further exemplify below is that the acuteness of different sensory signals can be comparable across modalities if they reflect an equally critical problem. When one problem or threat is greater than another, that is the dominating factor which gains control over the system. In this case, sensors that are typically perceived as being used to make predictions can in some contexts suddenly become of a more acute nature and other sensory systems which in the default situation are more critical threat signals can then almost be seen as becoming supportive. Importantly, the behaviour of the sensor hasn’t changed, only the importance and criticality of the problem it monitors.

\begin{displayquote}
Conjecture 1.8: A problem signal in or between cells is comparable to any other problem signal, regardless of origin, when in compatible chemical or electric form at the same location, and the acuteness of such a problem is measurable in the magnitude of the signal.
\end{displayquote}

It should be clear here that a problem signal can be transmitted through the body in many forms. It could be in a chemical signal form like glutamate, GABA, or some monoamine or peptide, through the neural system or through the bloodstream. In the neural system, it could be in an electrical form, binary as with an action potential or analog as a hyperpolarizing or depolarizing membrane voltage deviation from its resting value, i.e. set point. It could even be in the form of a chemical imbalance throughout the body, like an overabundance or shortage of calcium. This also points at how such problems are resolved, as a hyperpolarizing signal could be balanced with a depolarizing signal to reduce its effect in some parts of the body, whereas in other parts it could end up resulting in a muscle contraction, or an overabundance of calcium could be solved by opening channels which moves such calcium into intracellular buffer systems, whereas a calcium-deprived state could be resolved by releasing calcium from the same buffer system.

Here follows several additional examples from everyday life that support such prioritization schema being pervasive in brain operation.

\textbf{Airway obstruction:} A simple example is when there for one or other reason arises a partial airway obstruction, including just an excess amount of mucus in our respiratory airways, then that becomes an acute situation that we need to handle before we can complete anything else, even though these same sensors are active throughout daily life to control the pressure of the airway walls~\cite{Yu2005-oh}.

\textbf{Priorities in the newborn baby:} In newborn babies, a very large part of the waking time centers around nutritional status. When the baby is hungry, that becomes the highest priority of the baby’s brain. Similarly, when the end-part of the colon intestine fills up, this becomes the highest priority of the baby’s brain to somehow resolve that sensory signal. If they are similarly critical, the baby may even engage in both activities at the same time. Once these problems are both resolved, the baby may engage in other behaviors or return to sleep, as the brain is no longer preoccupied with those sensory inputs.

\textbf{Eye tear film:} When the tear film that normally covers our eyes gradually changes quality, so that it leads to refractive changes and deteriorated quality of vision, then blinking becomes a high behavioral priority~\cite{Kastelan2024-xh, Nosch2016-ql} regardless of what other task we are currently engaged in. For most animals and humans, tasks like these can become well-trained to the level of reflexive, to the point where the brain hardly realizes the problems are being resolved, and allows other behaviors to co-occur in the meantime, even if at the root those problems are highly acute.

\textbf{Vestibular control:} During ongoing whole-body movements such as locomotion or balancing on one leg, there may gradually arise a sense of losing balance, and then that becomes a top priority, meaning that we try to superimpose on the ongoing movement pattern compensatory muscle activation patterns that can help prevent us from falling. If one is engaged in a complex task during the balancing, the complex task can continue only up until the loss of balance becomes imminent~\cite{Woollacott2002-ec}, as also suggested in more general terms in gait control~\cite{Al-Yahya2011-iv, Schulleri2024-il}.

\textbf{Adaptations during Muscle Fatigue and Pain:} Similarly, during ongoing whole body movement, there may be ligament pain or muscle fatigue~\cite{Cowley2014-th, Mudie2016-yi} that gradually develops or is chronically or subchronically present. Such conditions have been shown to lead to different neural adaptations in the muscle activation patterns~\cite{Mudie2016-yi, Burdack2020-ta, Hodges2015-yi, Madeleine1999-oz, Madeleine2003-oo} in order to still be able to conduct the same type of ongoing movement, but to distribute the load to muscles so as the ‘bad’ sensor signals are minimized earlier. Acute pain appears to result in a quest for a motor solution reducing nociceptive sensor activation, while chronic pain can instead lead to diminished motor flexibility~\cite{Madeleine2008-dh}. Onset of muscle fatigue can result in an increase in variability between adjoining lower limb segments, which demonstrates the ability of the neuromotor system to adapt and maintain performance~\cite{Mudie2016-yi}. In response to muscle fatigue, we alter our biomechanical movement patterns in a way that specifically preserves the goal relevant features of task performance~\cite{Gates2008-jo} but minimize fatigue related sensory activity. These observations suggest that we constrain ourselves to those movement solutions that over time result in the lowest level of pain~\cite{Dean2013-dh} or, if such solutions cannot be found, simply reduce movements overall given allowance of the environment.

\textbf{Auditory pain:} Unlike the skin, muscles, ligaments and inner organ tissues, there are no known tissue damage sensors (‘nociceptors’) in the auditory sensory system of the cochlea. Nevertheless, we can perceive situations with high sound volume as unpleasant, and may for example raise our hands to our ears to protect them. An obvious potential source of pain signals are sensory afferents in the tympanic membrane, whose sensory afferents are for example likely the primary source of pain during middle ear infection that can sometimes even risk rupturing the tympanic membrane. But in addition, the cochlea itself has auditory afferents with different mechanical thresholds, where a higher mechanical threshold in the cochlea equals a higher sound volume. The arrangement that runs along the entire frequency range within the cochlea, with three rows of inner hair cells together with a single row of outer hair cells may well be reflective of a system designed to be able to signal varying degrees of acuteness in terms of sound volume (i.e. in line with the threshold-dependent activation as discussed above for muscle and skin sensors). In support of this notion, the auditory sensory afferents from the outer hair cells are often unmyelinated and they have comparatively a high threshold of activation~\cite{Flores2015-ah}. In other organ systems, unmyelinated sensory afferents are typically used for nociceptive signals~\cite{Amann2015-et, Mense1996-al}, i.e. when the mechanical or chemical load has been so high that there is actual tissue damage. Lack of myelination means that the afferents conduct their signals orders of magnitude slower into the CNS, which can be argued to be reflective of that their functional role is to essentially never be activated and that other signals of falling level of acuteness (threshold) should be used by the brain to prioritize behavior that prevents them from becoming activated. In this case, the unmyelinated afferents can be part of a sensory system designed so that the brain can prioritize behaviors that lowers the risk of mechanical load that could break components of the auditory system.

\subsection{Relationship to classical feedback control}

\begin{figure}
  \centering
  \includegraphics[width=\textwidth]{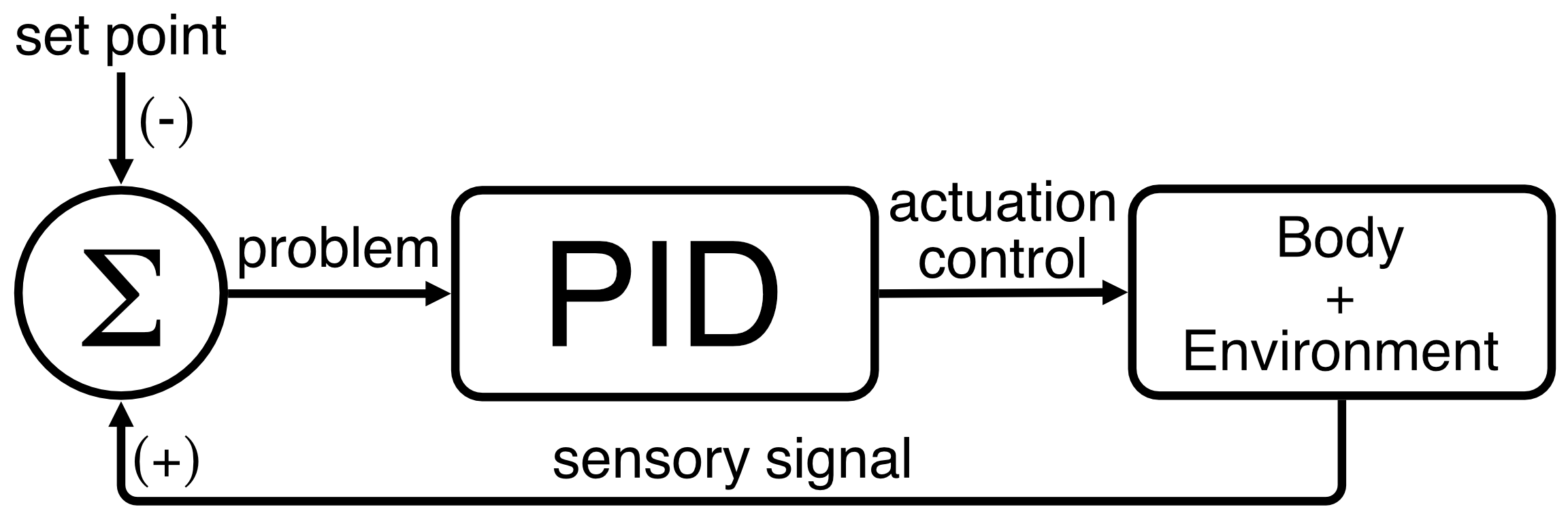}
  \caption{Principle of operation of a PID system in negative feedback control. A set point of the loop is compared to the actual current sensor signal. A deviation constitutes a ‘problem’ that needs to be fixed. The problem signal is fed through a converter that weighs in proportional (P) gain of the signal, an integration (I) of the signal and a derivation (D) of the signal to create an actuation control signal. The resulting actuation effect, i.e. depending on the actuation mechanisms that the cells have affordances for and the resulting impact on the environment, is sensed and fed back to the summator. In biology, the PID effects could be caused by various transformations, or representations of the problem signal, moreover in nested structures, that the cells and the networks of cells are capable of, as described in the Section 1.2.}
  \label{fig4}
\end{figure}

A central aspect of the proposed schema for sensory minimization is that each sensor signal (external or internal) has a potential intrinsic or homeostatic value, a target set point. Thereby, it shares important features with classical error-controlled regulation, such as in proportional–integral–derivative (PID) systems and similar low level feedback controllers (Figure~\ref{fig4}) many times previously implicated as a general mode of brain operation~\cite{Kawato1988-hk, Miall1993-yw, Wolpert1998-mv, Dean2010-bl}. Negative feedback control is one form of a PID system, which makes use of essential variables that are subtracted from a required value and the resulting control variable is applied to a regulator which can affect the controlled system in such a way that the essential variable approaches its set point.

In such systems, any defined parameter can be positive or negative and indicates its distance from a set point, but any parameter that is not attributable to a set point has no inherent function in the system and is excluded from its formulation.

In our formulation of sensors throughout this article, each sensor requires an attributable set point. In more advanced animals, there may be so many sensors signaling fine aspects of a particular input quality that each individual sensor may seem to not be directly relatable to sensor minimization, however, we purport that is only because that sensor tends to almost always be held sufficiently close to its set point to almost be considered informatory, even if a potential acute component is still present.

Hence, in this article we propose neuronal networks to have only sensory inputs which are increasing in relation to their measured deviation from a set point; i.e. an error for that parameter. These networks would exclude purely informative sensory inputs altogether unless they can be ascribed to a set point and error vector before being introduced to the network, and as observed even tactile sensors detecting shear forces, or retinal cones detecting light changes can be seen in this perspective. For example, the 50,000-100,000 tactile skin sensors in humans would then primarily be seen as ultrasensitive skin shear force sensors~\cite{Hayward2014-za, Jorntell2014-jn, Rongala2024-ad}, sensing risk of cell damage, or simply cell strain. The fact that the brain is able to use these lower threshold sensors for more intelligent behavior naturally doesn't contradict that sensing the risk of cell damage is the original reason they are there.

Whether too hot or too cold, lacking glucose or overabundance of glucose, receiving a lot of visual stimuli, stretching a muscle too far or applying too much force - these parameters are all problems that are minimizable by applying suitable output functions. We can sweat, release glucagon or insulin, focus our eyes and track movement, contract or relax our muscles - and through those functions improve our condition in the environment. Should all our inputs arrive at zero, any action could be said to be equally valid - and any sensor not at zero can be solved by executing any one or more actions that is expected to result in a negative activity slope for that sensor. If an input as we describe here can be inherently solved, then it is no longer purely information but an inherently optimizable parameter, leading the body toward homeostasis through learnable inputs.

\subsection{Cell learning applied to problem solving}
We next illustrate how the cell learning principles described above can be applied to solve problems in an example control task. It should be noted that this is a bare-bones model to illustrate how a neural or even a non-neural cell system could be constructed from our conjectures. While optimizations are surely possible they would bring an increase in the model complexity and are not needed for this illustration. For example inhibitory neurons, or cells whose output inhibits the activity of other cells, which could increase the behavioral repertoire of the system~\cite{spanne2015questioning}, are not included. The system implementations below share similarities with the neuronal circuitry of the spinal cord~\cite{jankowska1992interneuronal,kohler2022diversified}. But note that the systems illustrated below could in principle alternatively be implemented through adaptations in nested molecular cascade reactions (see Section 1.2) in an individual, non-neural cell. 

For this illustration, we chose the classical control problem of balancing an inverted pendulum (Figure \ref{fig5}A), in which a pole is balanced on a cart that moves along a rail. The controller can only act by applying forces to the cart with the goal of maintaining the pole in an upright balance. The simulation of the inverted pendulum was conducted in a third party physics simulator, MuJoCo \cite{todorov2012mujoco}. In the present implementation, there were two target set points for the control, the angle of the pole (in one plane only) and the position of the cart along the rail. Success was when these set points were maintained within a range close to zero for a sustained duration of several tens of seconds. To enable control, there were two main categories of sensors, those related to the angle of the pole and those related to the position of the cart. For each category of sensors, there was one sensor signal for each of the two directions and then in addition one sensor signal for the respective derivatives of those signals, yielding a total of eight sensors (Figure \ref{fig5}B). The sensor inputs, i.e. the outputs of the sensor cells, which in a neural system would correspond to its primary sensory afferents, were provided to two actuator cells ($M_{0}$, $M_{1}$; which could be said to correspond to alpha-motorneurons in the spinal cord). The actuator cells pushed the cart with a force to either the right or to the left, which depended on their relative activation levels (Figure \ref{fig5}B).  

\subsubsection{Implementation of the decentralized sensory learning process}
\begin{figure}
 \centering
\includegraphics[width=1\linewidth]{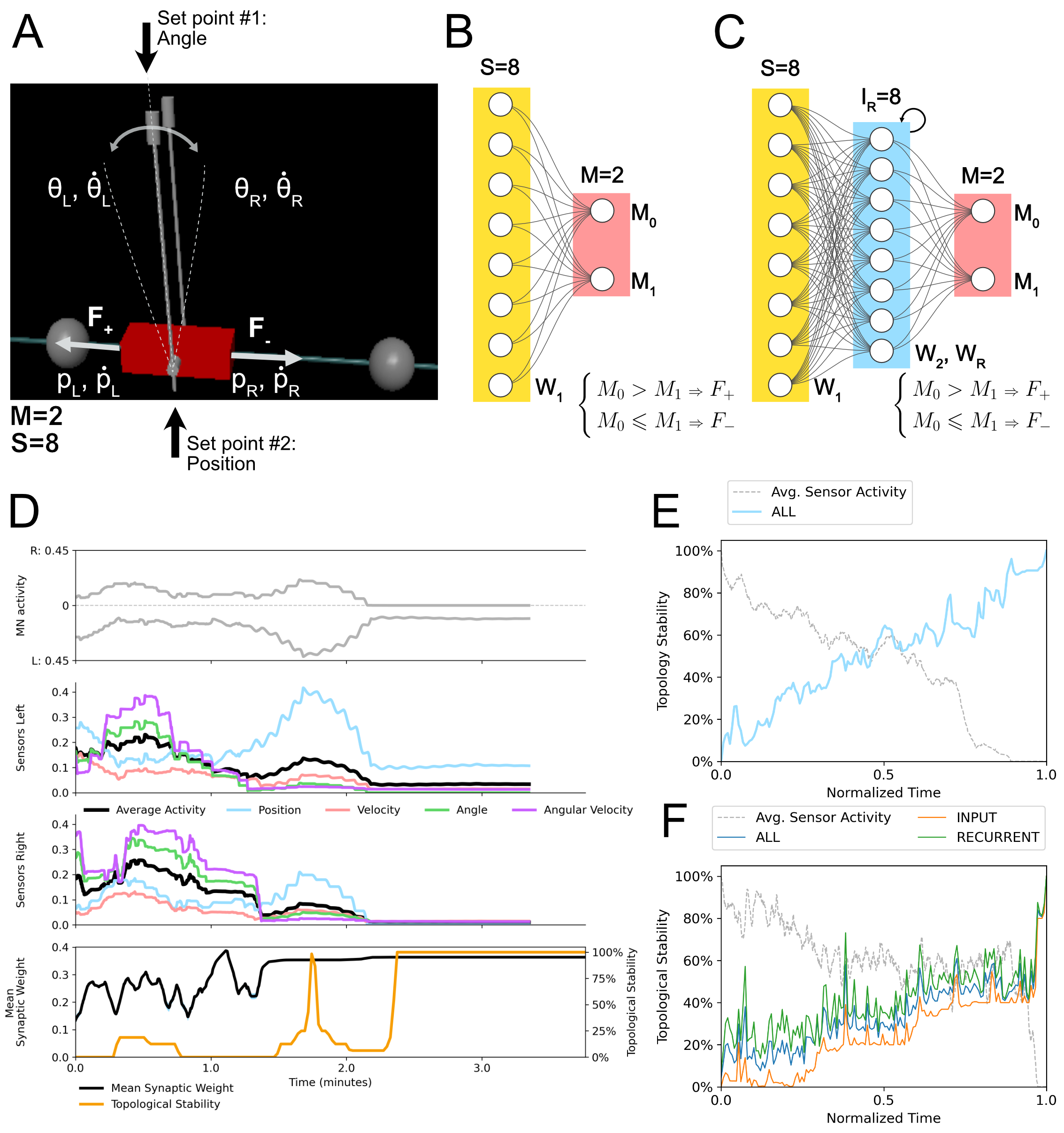}
\caption{Example implementation of the learning model based on the present conjectures. (A) The inverted pendulum implemented in the MuJoCo simulation environment~\cite{todorov2012mujoco}, with the 8 sensors (S=8), angle and position plus their respective derivatives, labeled by the two alternative directions, (L)eft and (R)ight. The figure also indicates the two output actions, F+ and F-, driven by the two output cells (M=2). (B) The simple cell system configuration, where the eight sensor neurons (yellow) project to the two output neurons (red) via weighted synapses ($W_{1}$). The action selection (F- or F+) were calculated from the activity of the two output neurons as indicated. (C) The larger cell system configuration with an intermediate layer (blue), which also featured recurrent synaptic connections between all of the neurons within the layer. (D) Example data from one complete learning process using the network in B. The top panel shows the running averages of the two motor neuron activities (30 second boxcar window). Middle two panels show the running averages for the sensor activities (colored lines), and the average total sensor activity (black line), split per direction. The bottom panel shows the running average of the synaptic weights ($\pm$3 standard deviations of the synaptic change per timestep; 5 second boxcar window), as well as the running average of the topological stability of the network (i.e. the inverse of the rate of synapse movements between the target neurons; 10 second boxcar window). (E) Running grand averages of the topological stability and the total sensor activity, respectively, for N=10 simulations using the network configuration in B. (F) Similar display as in (E) but for the larger network configuration (shown in C; N=5 simulations). The topological stability was illustrated separately for the input to the I-layer and the recurrent connections within the I-layer. Both panels E and F are normalized with respect to the time until reaching a stable solution.}
  \label{fig5}
\end{figure}

For each cell that received synaptic inputs, which in the first configuration (Figure \ref{fig5}B) only applied to the two output cells, the following set of equations describes how the synaptic inputs were adapted as a function of their activation and the output of the cell ('cells' will henceforth be referred to as 'neurons'):

\begin{equation}
    A = \sum_{i = 1}^{n}a_i . w_i 
    \label{eq1}
\end{equation}

\begin{equation}
    \Delta w_i = 
    \begin{cases}
            \eta, & \text{if } a_i > 2A \\
            -\eta, & \text{if } A > 1 \\
    0, & \text{if } a_i > 2A~\&~A > 1 
    \end{cases}
    \label{eq2}
\end{equation}


In addition to weight changes, the network was also adapting its structure over time. If a neuron had activity (A) greater than 1 at the same time as the input of an individual synapse ($a_i$) to that neuron was greater than 1, then that synaptic connection was disconnected and reinserted randomly to eligible recipient neurons (see Appendix 1). This is a specific adaptation to fulfill the conditions of the theory that \textit{a sensed problem must not be ignored or discarded until it has been resolved}. It follows that more than one synaptic connection between two specific neurons could exist. This feature was indeed also part of the random initialization of the network as described next.

In short, if the activity of the incoming synapse is higher than that of the receiver neuron, then strengthen the weight of the synapse. If the receiver neuron activity is too high, then reduce all synaptic weights. If the activity in both the incoming synapse and the receiver is high, then disconnect that synapse from the receiver and reinsert it on some eligible receiver neuron.

\subsubsection{The learning process for the simpler cell system configuration}
In the first example network (Figure~\ref{fig5}B), we provided the sensor signals as weighted synaptic inputs directly to the two motor neurons. Each sensory neuron  projected 25 synapses to these two eligible recipient neurons with initial randomization. Since there were only two motor neurons but eight sensory neurons there was a high degree of synaptic overlap from the beginning. The physics simulation of the inverse pendulum was started and the connectivity of the network was adapted according to the sensory minimization principle: if a synapse ($i$) is highly active ($a$) in the receiving neuron, then update the weight ($w_i$) of that sensor synapse. This will lead to a change in the output of the neuron ($A$) (Equation~\ref{eq1}), which aims to decrease its own sensor input signals over time (Equation~\ref{eq2}). Furthermore, by having two set points (both set to zero) we enabled this implementation to demonstrate the principle of sensory prioritization — the learning rule prioritizes adaptations of the most active sensor signals within each neuron (Equation~\ref{eq2}). Adaptations consisted in weight increases for highly active sensors, or in weight decreases if the output activity of the neuron was too high (Equation~\ref{eq2}), whereas if both the input activity and the output activity was high then the synapse was removed from that neuron and randomly assigned to another eligible neuron (including the same cell; see Appendix 1). The latter was hence an adaptation to the case where a solution could not be found with the current anatomical synapse connectivity — then the synapses with the highest activity was instead randomly re-routed. Once a sensor signal was brought down, other sensor signals would be of higher magnitude and then the neuron focused its learning to bring down also those sensor signals. This could lead to that other sensor signals rose in magnitude as a consequence of the new adaptations, resulting in a natural staging of how the solution evolved. Evidence of such phased problem solving of specific sensor inputs can be seen in the plot of the sensor signals (in Figure~\ref{fig5}D). The next plot (Figure~\ref{fig5}E) illustrates that the sum of all sensory inputs gradually went down while the network topology stabilized in parallel, until the solution was found and no further changes occurred. Across different random initializations, in some cases the later stages of adaptation could fail, which could lead to previously solved sub-problems reoccurring, and then the solution discovery process will move back and forth until a stable solution can be achieved.

\subsubsection{The learning process for the larger cell system configuration}
Using the same learning principle, we also tested a system configured with another layer of neurons between the sensors and the actuators, where all the neurons within this intermediary layer were potentially recurrently connected to each other (Figure~\ref{fig5}C; like the sensor neurons, each intermediary neuron issued 25 synapses that were initially randomly distributed to motorneurons and other intermediary neurons). The sensory input layer only provided input to the intermediate layer (i.e. this cell population would be the equivalent of spinal interneurons~\cite{jankowska1992interneuronal,kohler2022diversified}), and all the cells in the intermediate layer innervated the two actuator cells. This system adapted in a similar manner as the simpler configuration, i.e. by achieving a gradually decreasing sensory input in parallel with a gradually stabilizing synaptic topology, but the solution was more stepwise led by stabilization of the recurrent connections within the intermediary population (Figure~\ref{fig5}F). In this configuration we hence illustrated the principle that cells will propagate their unresolved problems to other cells (Conjectures 1.4-1.8), which may thus indirectly help to reduce the sensory input to that cell \textit{by utilizing downstream cells as remote enactors to resolve the problem sensed by that cell}. 

For both of these cell systems (Figure~\ref{fig5}B,C), the synaptic weights evolved until a stability in the network topology was reached (Supplementary Figures 1-4). It is noteworthy that even with this bare bones model, different cell system configurations were able to solve the same problem.

\subsubsection{Generalization of the learning model}
\begin{figure}
    \centering
    \includegraphics[width=0.5\linewidth]{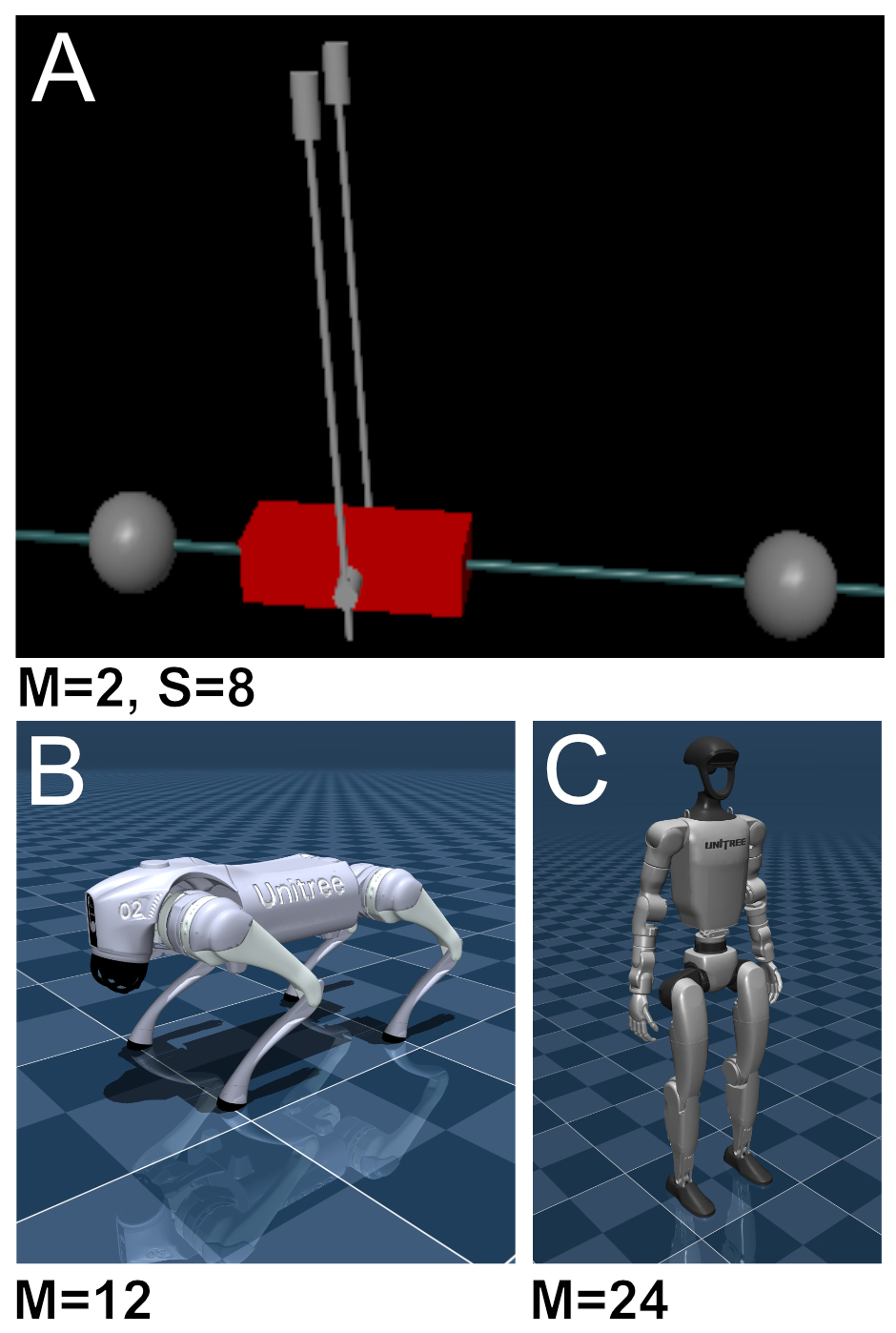}
    \caption{Different environments used as contextual realization for 
    more advanced implementations of the learning model, showcasing the potential of the current conjectures. (A) Inverted pendulum, same environment as above, with 2 motor outputs and 8 sensors. (B) Unitree Go2 quadruped robot, illustrated learning process in real world in \href{https://drive.google.com/file/d/1xZLYs8OKsyzQgVg1SJbtV8uby5GB8bEe/view?usp=sharing}{video 1} for 12 motor outputs. (C) Unitree G1 bipedal robot, illustrated learning process for 24 motor outputs in simulation~\cite{todorov2012mujoco} in  \href{https://drive.google.com/file/d/1JabBySbRVh77s0ZVaij-FBDSYlVfeSfR/view?usp=sharing}{video 2}. }
    \label{fig6}
\end{figure}

The simplicity of the problem context, the inverted pendulum, provides a limited evaluation of the breadth of the applicability of the learning model. However, based on the same foundational conjectures, adaptations of this learning model can be implemented to also solve very different problems related to systems with more biological-like anatomical structures, e.g. a quadruped and a humanoid robot (Figure~\ref{fig6}). In these cases we employed non-disclosed proprietary implementations. We illustrate the macroscopic learning processes achieved for the quadruped (real-world learning) and the biped humanoid (simulation in MuJoCo~\cite{todorov2012mujoco}) in \href{https://drive.google.com/file/d/1xZLYs8OKsyzQgVg1SJbtV8uby5GB8bEe/view?usp=sharing}{video 1} and \href{https://drive.google.com/file/d/1JabBySbRVh77s0ZVaij-FBDSYlVfeSfR/view?usp=sharing}{video 2} respectively.

\subsection{Relation to Learning Frameworks and Plasticity}

It is our hope that the framework we provide here can provide new insights into leading neuroscience theories and experimental observations on biological learning, for example why Hebbian learning and spike timing dependent plasticity are inferrable in experiments, while networks over time still arrive at highly stable homeostatic states. It should be noted that observations that unicellular organisms learn dates back more than 100 years (reviewed by Gershman~\textit{et. al.}~\cite{Gershman2021-xq}) and the idea that learning in the brain can be seen as a result of the need to keep the individual cells in homeostatic balance was proposed before~\cite{Levin2019-nv}. We build on and extend those ideas towards more specific mechanistic explanations through formalizing what is being sensed, how it is being sensed and prioritized, and the passing of locally unresolved problems as internally created sensor signals to more suitable subspaces.

The focus on sensors in this framework may appear to share some resemblance with the widely researched conceptual framework of sensory prediction errors~\cite{Shadmehr2010-tm, Schlerf2012-mo, Auksztulewicz2017-nr, Tsay2022-ok} or adaptive filters~\cite{Dean2010-bl, Nilsson2023-nv} in neuroscience. We believe our conjectures provide a tool to understand how results interpreted within the sensory prediction error theory can arise as a result of the development in the neural network of the brain being driven by the sensory minimization principle, which can allow such an apparent predictive system to arise within the brain without supervision.

In recent years, the detectability of reward signals within the brain has been brought to light~\cite{Cohen2012-hh, Yamagata2015-nk, Li2016-io, Cox2019-gk, Jeong2022-uk}, and through our conjectures we show that these signals would be expected by-products of the underlying principle where changes in activity from solving a nonlocal problem will be detectable at many points throughout a decentralized network and can be interpreted as a reward corresponding to the magnitude of a solved problem. With this in mind we also see a possibility that structuring sensory input as problems can be a sufficient criteria for decentralized learning from reward to be realised.

In Artificial Intelligence, input is typically in the form of structured data, to be transformed with the intent of extracting more valuable data. By supplying large amounts of such input to these networks and utilizing training to translate that data into valuable information for a target task, the field of deep learning builds intelligent systems. We want to applaud the impressive results in that field. However, here we show through the presented conjectures that biological organisms very likely do not share the same process and most probably cannot intrinsically value raw data in the form that deep learning makes use of it. Instead, biological organisms sense only solvable, value-estimable problems that need to be surmounted to ensure survival, and this understanding of when there is a need to adapt is what allows those organisms to learn in the first place.

\section{Conclusion}

We provided a theory that explained how the first unicellular life forms could survive by propagating problem signals from various cellular sensors across the cell to solve problems in order of priority.

We then presented how that same theory directly translates into communication between cells in a multicellular organism, enabling cells to have more specialized functionality and solve problems at larger scales more effectively.

After that, we discussed how evolution could pressure the development of electrical signaling to allow the same problem propagation used by unicellular and multicellular organisms to function across larger distances and diverse sensors throughout a body - the emergence of the nervous system.

Next, we inferred that the same principles that arose at the unicellular level are in use by neural tissues of animals even today to solve problems across all modalities, whether internal or external.

Furthermore, we presented a bare bones implementation based on the conjectures that was able to solve a classic control problem and demonstrated how other implementations of the conjectures have been applied to solve the control of quadruped and biped systems in real-world and in simulation, respectively. 

We discussed how our theoretical framework can act as a unifying principle to support many experimental neuroscientific observations, and also why these sensory signals are the only necessary value signals for biological life and the brain to function and learn, as well as how this differs from modern artificial intelligence.

Through this whole article, we have provided a set of conjectures that can explain the adaptations that occur across life at all scales and timeframes to enable learning, survival and proliferation.

\section{Future Work}

We believe the conjectures we propose here, should they prove to apply, can open up venues for new interpretations of various biological data collected over many years, new ways to approach mental health, and new ways to approach and understand intelligence.

We hope the wider neuroscientific and cellular biology communities can provide further insights as to how practical experiments could be set up to help prove or disprove our proposed conjectures in the respective fields.

\section{Conflict Of Interest}
Linus M\aa rtensson, Jonas Enander, Udaya B. Rongala and Henrik J\"orntell are founders of IntuiCell AB, a company dedicated to providing new solutions for artificial intelligence based on novel findings in the field of neuroscience.

This work was made possible by research performed at IntuiCell AB.

\bibliography{ReferencesSensing}

\section{Appendix 1}

\begin{algorithm}
\caption{The relationship between synaptic learning and neuron activity}\label{alg:cap}
\begin{algorithmic}
\Ensure $ 0 < W < 1$

\If{$S$ is connected} \Comment{for a given synapse $S$ between source neuron $i$ and receiving neuron $j$} 
    
    \State $A_j \gets a_x . w_x$ \Comment{$A_j$ summed activity of receiving neuron.}
    
    \If{$A_i > 2*A_j$} \Comment{synaptic potentiation}
        \State $\Delta W += LR$ \Comment{$LR$ learning rate} 
    \EndIf
    
    \If{$A_j > 1$} \Comment{synaptic depression}
        \State $\Delta W -= LR$
    \EndIf

    \If{$A_i > 1~\&~A_j > 1$} \Comment{$A_i$ activity of source neuron}
        \State $S \gets disconnected$
    \EndIf
    
\EndIf
\\

\If{$S$ is disconnected} \Comment{procedure for reinserting disconnected synapse}
    \State $S \gets P(X,L)$ \Comment{$X$ is random neuron index within a possible cell layer $L$}
    \State $W = 0$
\EndIf

    
    
\end{algorithmic}
\end{algorithm}

\newpage

\setcounter{figure}{0}
\makeatletter 
\renewcommand{\figurename}{Supplementary Figure}
\makeatother

\section*{Supplementary Figures}

\newpage

\begin{figure}
    \centering
    \includegraphics[width=1\linewidth]{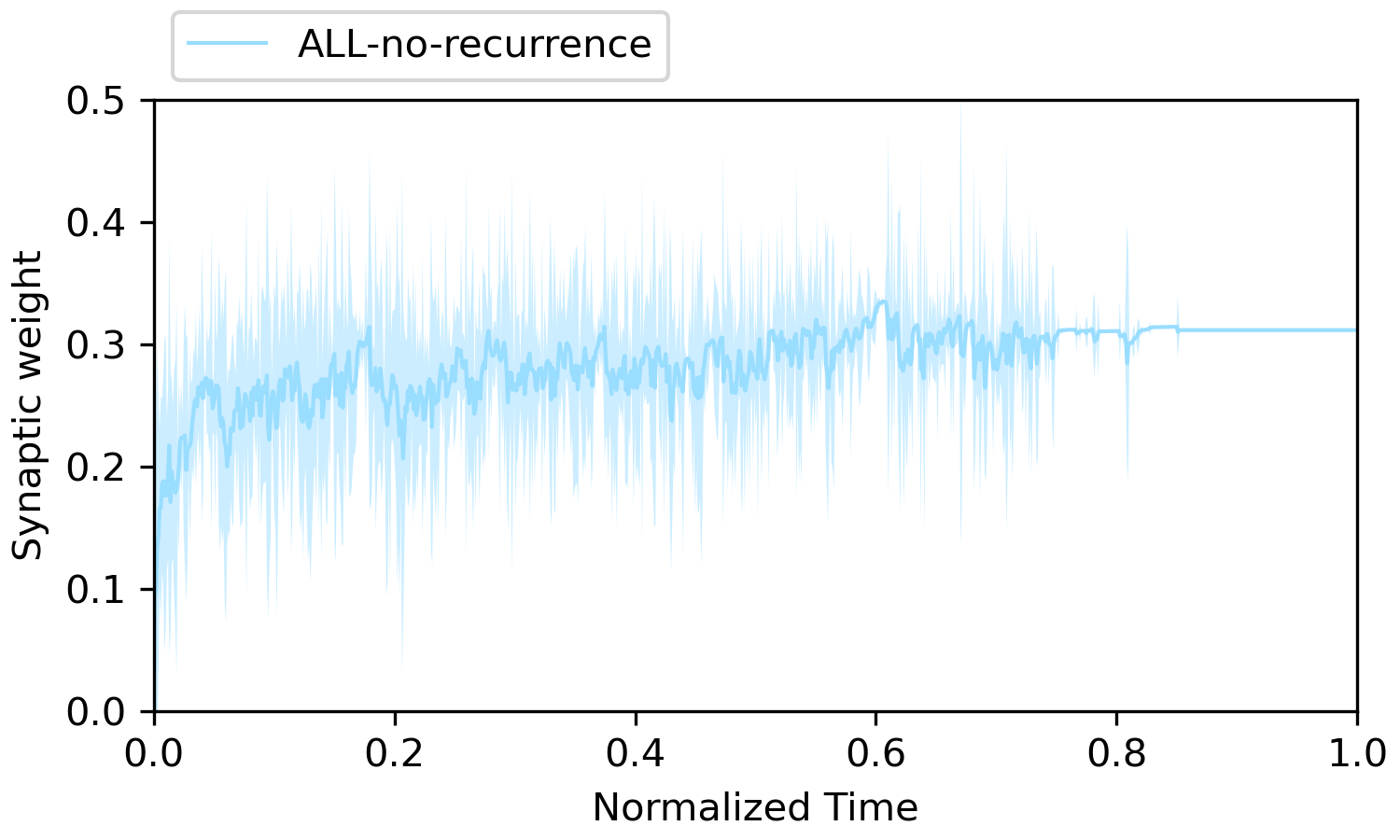}
    \caption{Grand average synaptic weight over normalized time for the simple cell system. Less saturated blue indicates ±2 standard deviations (s.d.s) for the difference between timesteps.}
    \label{figS1}
\end{figure}

\newpage

\begin{figure}
    \centering
    \includegraphics[width=1\linewidth]{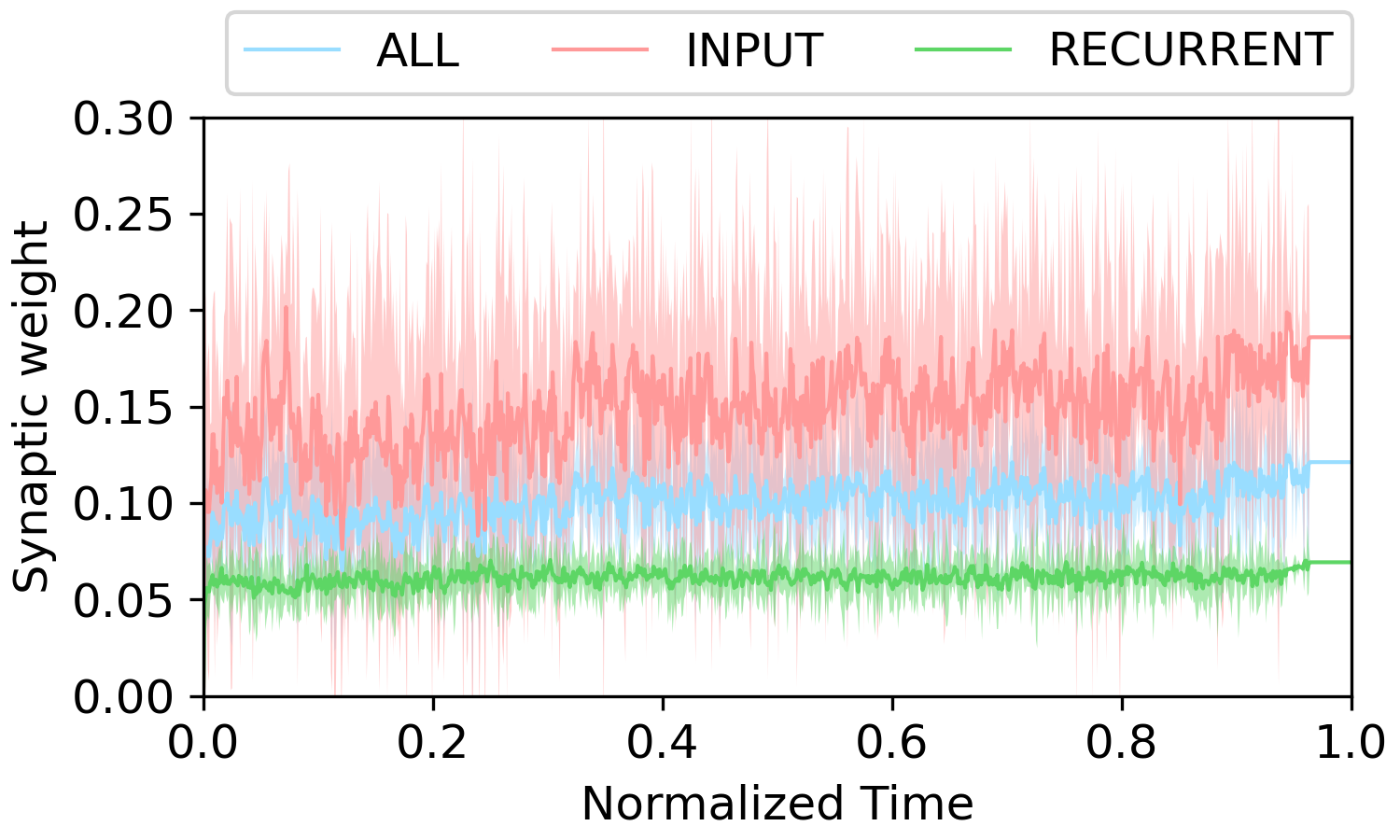}
    \caption{Grand average synaptic weights over normalized time for the larger cell system, shown for connections from the Input layer (red), from and within the Recurrent layer (green), and all of the connections (blue). Less saturated colors indicate ±2 s.d.s for the differences between timesteps.}
    \label{figS2}
\end{figure}

\newpage

\begin{figure}
    \centering
    \includegraphics[width=1\linewidth]{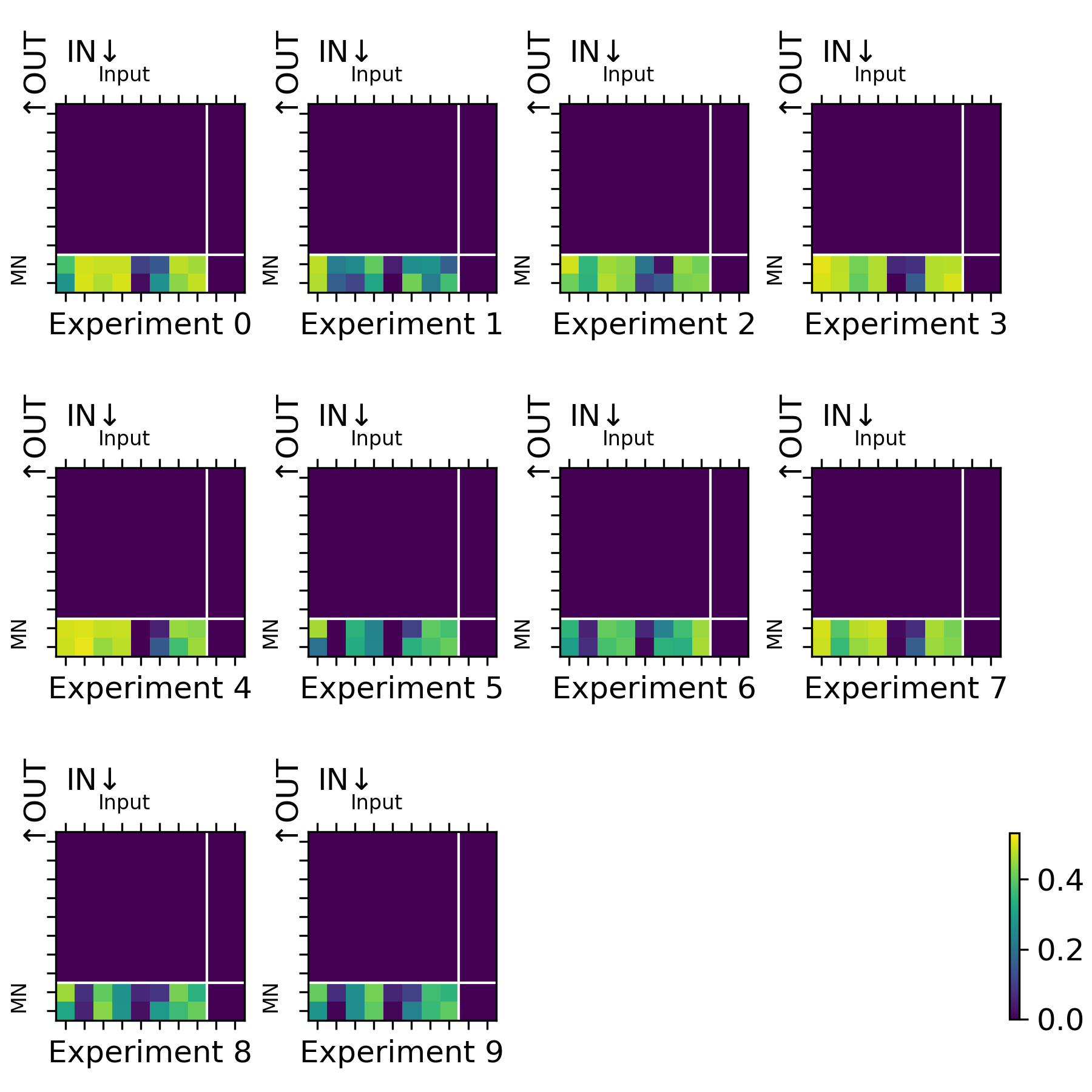}
    \caption{Matrices of the relative connectivity strengths (calibrations in the color bar) between individual sensors and motor neurons (MN) at the end of the simulation for the 10 different random network initializations for the simple cell system.}
    \label{figS3}
\end{figure}

\newpage

\begin{figure}
    \centering
    \includegraphics[width=1\linewidth]{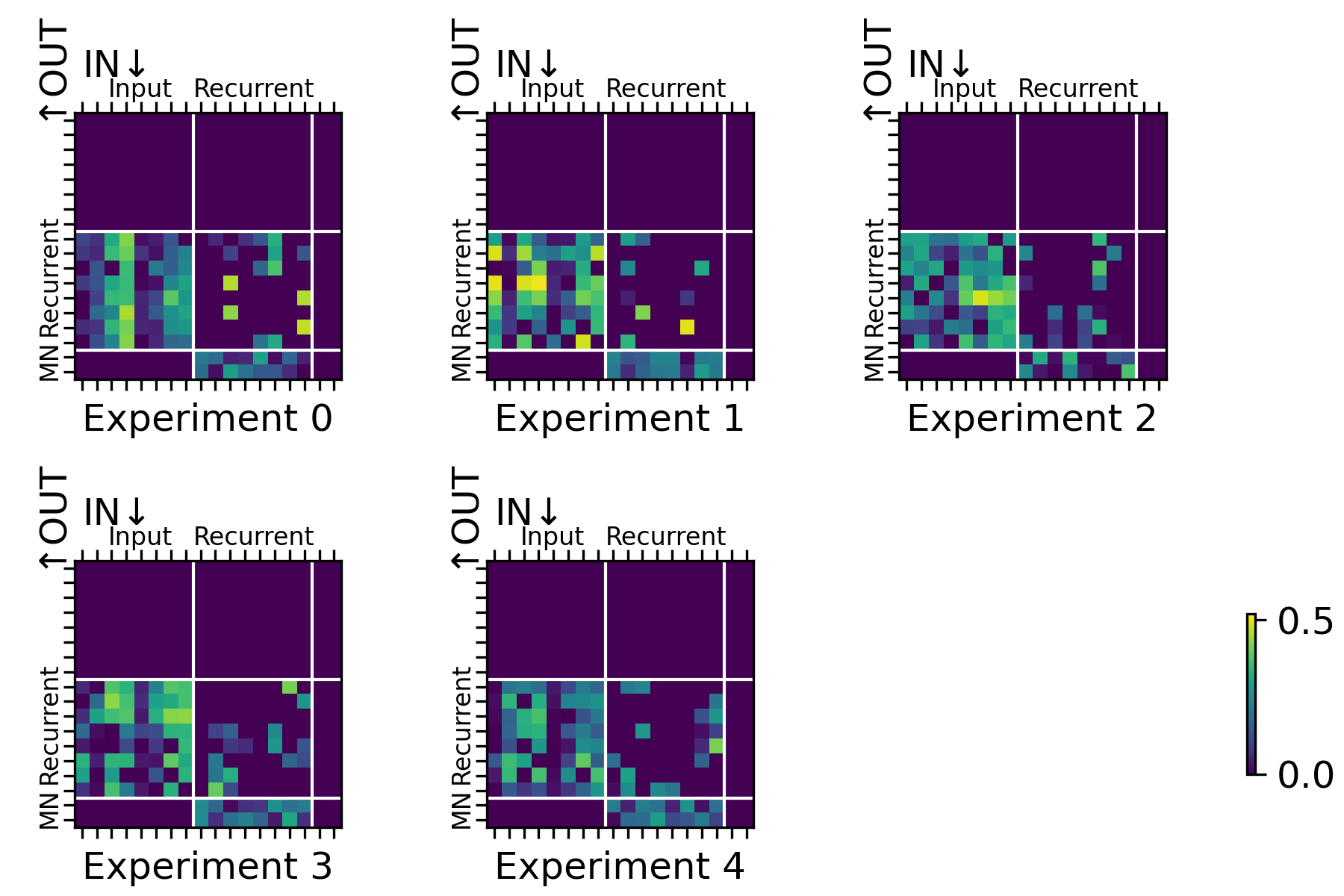}
    \caption{Matrices of the relative connectivity strengths (calibrations in the color bar) between individual sensors and intermediary neurons (‘recurrent’), within the intermediary layer, and from the intermediary layer to the motor neurons (MN) at the end of the simulation for the 5 different random network initializations for the larger cell system. The relative simplicity of the problem context, the inverted pendulum, is reflected in the comparatively sparse utilization of the recurrent-recurrent connectivity implying a low utility of latent space computation while simultaneously arriving at a stable problem solving solution.}
    \label{figS4}
\end{figure}

\end{document}